\documentclass[12pt]{article}
\usepackage {graphicx}

\newcommand{\prd}[3]{Phys. Rev.~{\bf D~#1} (#3) #2}
\newcommand{\plb}[3]{Phys. Lett.~{\bf B#1} (#3) #2}
\newcommand{\npb}[3]{Nucl. Phys.~{\bf B#1} (#3) #2}
\newcommand{\cmp}[3]{Comm. Math. Phys.~{\bf #1} (#3) #2}
\newcommand{\cpc}[3]{Comput. Phys. Commun.~{\bf #1}: (#3) #2}

\newcommand{\m}[1]{\mathrm{\bf{#1}}}
\newcommand{\mathin}[2]{\mathrm{\:\:In[#1]}:=\,\,\m{#2}}
\newcommand{\mathout}[2]{\mathrm{Out[#1]}=\,\,\mathrm{#2}}
\newcommand{\timing}[1]{&&\nonumber\mathrm{Timing\::=\,\, {\bf #1}~seconds}}

\newcounter{bla}

\begin{document}

\begin{center}
\Large{\noindent{\bf Package for Calculations and Simplifications
of Expressions with Dirac Matrixes (MatrixExp)}}\end{center}
\vspace{1cm}
\begin{center}
{V. A. Poghosyan{\footnote {\hspace{-0cm}email:
vpoghos@server.physdep.r.am\\The package is available at
http://www.yerphi.am/matrixexp.html}}\\Yerevan Physics Institute, 2
Alikhanyan Br., 375036 Yerevan, Armenia},
\end{center}
\vspace{0.5cm}
\begin{abstract}
This paper describes a package for calculations of expressions
with Dirac matrixes. Advantages to existing similar packages are
described. MatrixExp package is intended for simplification of
complex expressions involving $\gamma$-matrixes, providing such
tools as automatic Feynman parameterization, integration in
$d$-dimensional space, sorting and grouping of results in a given
order. Also, in comparison with existing similar package Tracer,
presented package MatrixExp has more enhanced input possibility.
User-available functions of MatrixExp package are described in
detail. Also an example of calculation of Feynman diagram for
process $b\to s\gamma g$ with application of functions of
MatrixExp package is presented.

\vspace{0.5cm}
\begin{flushleft}

\end{flushleft}
\end{abstract}

\vspace{0.5cm}
\newpage
{\bf PROGRAM SUMMARY} \vspace{0.3cm}

\begin{small}
\noindent {{\em Keywords:} Computer algebra, Particle Physics,
Computing, Quantum Field Theory, High energy physics, Calculation,
Gamma-algebra,Dimensional Regularization\\
{\em Classification:} 11.1 General, High Energy Physics and Computing, 4.4 Feynman diagrams\\
{\em Nature of problem:}\\
Feynman diagram calculation, simplification of expressions with
$\gamma$-matrixes
   \\
{\em Restrictions:}\\
MatrixExp package works only with single line of expressions
(G[l1,...]), in contrary with Tracer package which works with
multiple lines, i.e. the following is possible in Tracer, but not
in MatrixExp : G[l1,...]**G[l2,...]**G[l3,...]..., which will
return the result of G[l1,...]**G[l1,...]**G[l1,...]...
   \\
{\em Running time:}\\
Seconds for expressions with several different $\gamma$-matrixes
on Pentium IV 1.8GHz and of the order of a minute on Pentium II
233MHz. Calculation times rise with the number of matrixes. }

\end{small}

\newpage


\hspace{1pc}
{\bf LONG WRITE-UP}

\section{Introduction}

Modern calculations of radiative corrections in Relativistic
Quantum Field Theory, in particular within the limits of Standard
Model, lead to calculations of one-, two- and higher-loop Feynman
diagrams. Considering, that calculations of two-loop and
furthermore multi-loop diagrams are labor-consuming enough and
volumetric, and also that the number of possible diagrams to be
evaluated is great, now are widely applied various packages of
symbolic manipulations. In particular, Tracer package
\cite{Tracer} working in MATHEMATICA \cite{Mathematica}
environment is known, allowing to operate with expressions of
$\gamma$-matrixes in $d$-dimensional space. Following the
developers of Tracer package we chose MATHEMATICA environment for
the MatrixExp package, since MATHEMATICA is widely used by
researchers and is easily programmable, and also, which is more
important, unlike similar package MAPLE \cite{Maple}, MATHEMATICA
allows to program in ``rule-based'' style: a special style of
programming convenient for implementation (application) of
mathematical knowledge into the program (package). From the
technical point of view, during calculations of radiative
corrections it is necessary to get rid from arising intermediate
singularities, inherent to quantum corrections, usually by the
method of dimensional regularization; hence, it is necessary to
perform the calculations in $d$-dimensional space. The MatrixExp
package uses the so-called Naive Dimensional Regularization scheme
(NDR). As it is known, the given scheme leads to some algebraic
inconsistencies in $d$-dimensional space
\cite{Breiten:1977,Bonneau:1981}. However, as it has been shown in
Ref. \cite{Buras:1990}, in many calculations which don't involve
traces with $\gamma_5$ matrix, NDR scheme gives correct result. A
problem connected with definition of $\gamma_5$ matrix in
$d$-dimensional space is solved in the t'Hooft-Veltman
regularization scheme \cite{tHooft:1972,Akyea:1973}. In future,
this scheme will also be included in the MatrixExp package.

\section{Tracer package}

Tracer package is written basically in ``rule-based'' style which
is presented in \cite{Tracer}. We shall briefly describe this
style since our developed program also is written in this style.
Unlike the procedural style of programming, where the program is
executed consistently from the first command up to the final,
realizing the algorithm for solving the task in view,
``rule-based'' style is setting up a collection of transformation
rules which just represent mathematical rules of transformations
and properties of the given objects. The decision of a task in
view in ``rule-based'' style is reduced to the following:

\begin{enumerate}
\item{Setting up of transformation rules (which are applied when
expression matches the pattern on the left-hand side (lhs) of the
rule, and return the right-hand side (rhs) of the rule)}
\item{Use of basic rules built-in in MATHEMATICA}
\item{Consecutive application of corresponding rules
(automatically carried out by MATHEMATICA)}
\item{Last received expression which does not match to any pattern
is the received result.}
\end{enumerate}

Schematically the solution of a problem in ``rule-based'' style is
presented on Fig.~1.

\begin{figure}[]
\vspace{0cm} \hspace{0cm}
\includegraphics[width=14cm,height=13cm]{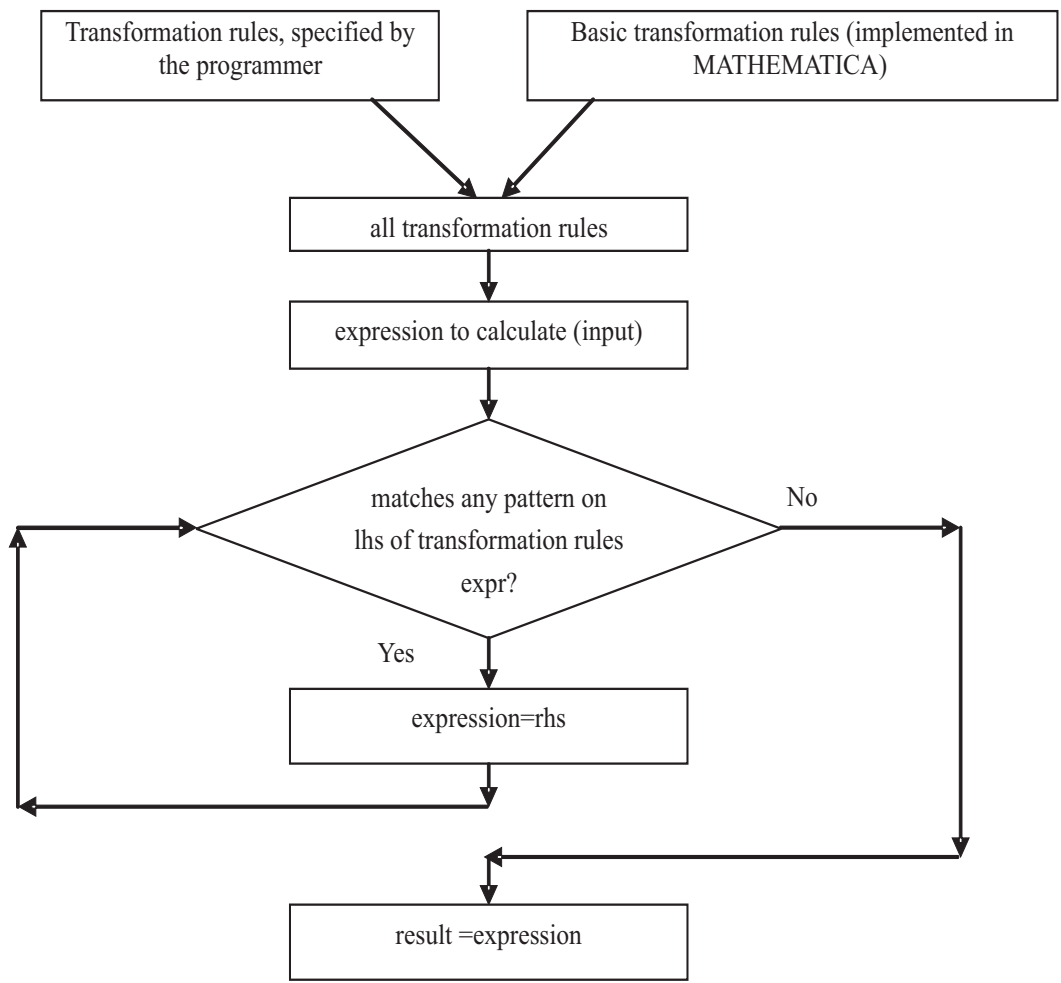}
\vspace{0cm} \caption[]{Scheme of problem solution in a
``rule-based'' style.}\vspace{0cm}
\end{figure}

The solution of a problem in this style greatly simplifies the
programming and does clearer the program code since allows to not
supervise all stream of the information during the solution of a
problem (at calculation of diagrams very bulky and complex
expressions arise), but by means of setting up of all
transformation rules (which are the initial mathematical rules,
i.e. are clear and natural to the user) directly solves the
problem in view (in our case - will transform initial expression
to a convenient form, carrying out all necessary transformations).
To illustrate the above mentioned, codes for differentiation in
procedural style and in ``rule-based'' style are presented in
Table~\ref{scheme2}.

\begin{table}[h]
\begin{flushright}
\end{flushright}
\hrule \vspace{0.2cm}
\begin{tabbing}
 {\bf a. Procedural style}\\
 dif\=f[y\_\,, x\_\,] :=\\
 \>Blo\=ck[\{n\}, \+\+\\
 If[T\=rueQ[Length[y]==2 \&\& y[[0]]==Power \&\& y[[1]]==x],\+\\
 n=y[[2]]; n*x\^\,(n-1),\\
 If[T\=rueQ[y==x], 1,\+\\
 If[TrueQ[Length[y==1] \&\& y[[0]]==Log \&\& y[[1]]==x], 1/x]]];
\end{tabbing}
{\bf b. ``rule-based'' style}\\
diff[x\_\^\,n\_\,, x\_\,] := n*x\^\,(n-1);\\
diff[Log[x\_\,], x\_\,]:= 1/x;\\
\vspace{-0.2cm} \hrule \vspace{0.2cm}
 \caption{Codes for
derivative calculation (cases of power function and logarithmic
function) in procedural {\bf(a)} and ``rule-based'' {\bf(b)}
styles.}\label{scheme2}
\end{table}

Note, that if we try to define differentiation for
all\footnote{e.g. for 5 or 10 more functons} other functions as
well, then the code in procedural style would get very
complicated, including special recursive codes for mixed
functions, code for parsing expression of functions etc., and in
the case of ``rule-based'' programming, one has just to add all
the definitions, rules for differentiation of a sum and product of
functions and rules for differentiation of $f(g(x))$.

Tracer package realizes $\gamma$-algebra in ``rule-based'' style
and calculates a trace of strings of $\gamma$-matrixes
\cite{Tracer}. Tracer accepts following types of data in a
corresponding format:

\begin{enumerate}

\item{a momentum or a sum of momenta $p$, $p-q$, $(p+k1-k2)$, which
correspond to $\hat p$, $\hat p-\hat q$ and $\hat p+\hat k_1-\hat
k_2$}
\item{an open index of a gamma matrix, which is denoted by a symbol in
curly brackets $\{\mathrm{alpha}\}$ corresponding to
$\gamma_\alpha$}
\item{a linear combination of the unit matrix in the gamma algebra $U$ and
the matrix $\gamma_5$ denoted by $G5$, with scalar coefficients}
\item{a linear combination of a momentum $p$ and a mass term $m U$.}
\end{enumerate}

For convenience, we use the same input-format for MatrixExp
package\footnote{except points (3) and (4). There are special
symbols for $(1\pm\gamma_5)/2$ defined as $R$ and $L$
correspondingly. $\gamma_5$ also may be used in the same way, i.e.
$G5$. For point (4) see sections 3 and 4.}, for users already
familiarized using Tracer package. Also, this way one may directly
use results achieved by Tracer as input for MatrixExp and vice
versa.

\section{MatrixExp Package}

When calculating diagrams, usually so-called Feynman
parameterization \cite{Peskin:1996} is applied, therefore, in
initial expression instead of momenta $p$ arises their linear
combinations, for example $p+u_1*k+u_2*r...$ and so the direct
application of Tracer package is already impossible, i.e. it is
necessary to segment (partition) the initial expression to parts,
not containing such combinations of momenta $(p, k, r...)$ and
scalars $(u_1, u_2...)$.

Developed new package MatrixExp presented in this paper, allows
entering data without additional transformations.

Input data for calculation by Tracer and MatrixExp packages
correspondingly (b) and (c), is presented on Table~\ref{scheme3}.

\begin{table}[h]
\begin{flushright}
\vspace{0cm}
\end{flushright}
\hrule \vspace{0.3cm}
{\bf (a) Expression to be calculated}\\
$\langle
s|\hat{p}_s,\gamma_\alpha,\hat{r},\hat{p}-m_b,\hat{k}+\hat{r},\gamma_\alpha|b\rangle,
\;\hat{r}\to\hat{r}-u_1\hat{p}+u_2\hat{p}_s $\vspace{0.2cm}\\
{\bf (b) Input for Tracer package}\\
a11 = G[l1, ps, \{alpha\}, r, p-U mb, k+r, \{alpha\}];\\
a12 = -u1*G[l1, ps, \{alpha\}, k, p-U mb, k+r, \{alpha\}];\\
a13 = u2*G[l1, ps, \{alpha\}, ps, p-U mb, k+r, \{alpha\}];\\
a21 = -u1*G[l1, ps, \{alpha\}, r, p-U mb, k-p, \{alpha\}];\\
a22 = u1\^\,2*G[l1, ps, \{alpha\}, k, p-U mb, k-p, \{alpha\}];\\
a23 = -u1*u2*G[l1, ps, \{alpha\}, ps, p-U mb, k-p, \{alpha\}];\\
a31 = u2*G[l1, ps, \{alpha\}, r, p-U mb, k+ps, \{alpha\}];\\
a32 = -u2*u1*G[l1, ps, \{alpha\}, k, p-U mb, k+ps, \{alpha\}];\\
a33 = u2\^\,2*G[l1, ps, \{alpha\}, ps, p-U mb, k+ps, \{alpha\}];\\
result = a11+a12+a13+a21+a22+a23+a31+a32+a33;\vspace{0.2cm}\\
{\bf (c) Input for MatrixExp package}\\
Vsetscalars[u1, u2, mb];\\
a = G[l1, ps, \{alpha\}, r-u1*p+u2*ps, p-mb, k+r-u1*p+u2*ps,
\{alpha\}];\\
result = Vcalc[a];\\
\vspace{-0.2cm} \hrule \vspace{0.2cm} \caption{Calculation of
expression {\bf(a)} using Tracer package {\bf(b)} (as we see, one
needs to segment the initial expression into parts, and if the
expression is more complex, then the calculation becomes very
complicated) and using MatrixExp package {\bf(c)} (Function
Vsetscalars[] declares $u1$, $u2$ and $mb$ as scalars and then
function Vcalc automatically performs all corresponding
operations).}\label{scheme3}
\end{table}

Thus, MatrixExp package allows working with complex expressions,
and there is no need for additional manual transformations of
initial input expression. Besides, since it is very often applied
Feynman parameterization, with the subsequent $d$-dimensional
integration, this opportunity is also included in the MatrixExp
package, i.e. it is possible to set both automatic Feynman
parameterization, and automatic integration. The package is
intended not only for calculation of traces of matrixes, but also
for calculations and simplifications (sorting and grouping) of
complex expressions received as a result of previous calculations
or as output from other packages. During calculations MatrixExp
package operates with scalar products (additional conditions
imposed on scalar products can be set via appropriate
user-function of MatrixExp (see next section)), combines and
recombines momenta according to $\hat p = \gamma_\alpha p_\alpha$
for performing calculations, integrations and grouping results
(for example, $p_\alpha\hat{q}k_\beta(u_1 r_\alpha +
\gamma_\alpha)\gamma_\beta...\to\hat{q}(u_1(p.r)+\hat{p})\hat{k}$...,
etc.).

MatrixExp package includes such enhancements as definition of
scalars, automatic sorting by the given mask (order), automatic
Feynman parameterization (using previously declared scalar
symbols), grouping of sorted results, automatic $d$-dimensional
integration over specified momenta, automatic rearrangement of
"edge"-momenta ($p_b$ and $p_s$) and their replacement by
corresponding masses according to Dirac equation ($\langle
s|p_s=m_s\langle s|$ and $p_b|b\rangle=m_b|b\rangle$), operations
with scalar products, etc.

\section{User functions of MatrixExp package}

The detailed description of MatrixExp functions and their syntax
is included in the package and is accessible via standard method
in MATHEMATICA, i.e. $\m{`~?function\_name~'}$. Call
$\m{`~Vhelp[\,]~'}$ to receive the list of all functions defined
in the package.

Here we present basic functions and their purpose.

$\m{Vsetscalars[u1, u2...]}$ - declares symbols $\m{u1, u2...}$ as
scalars. By default the package declares as scalars the following
symbols: $\m{u1... u9}$ since they are necessary to perform
Feynman parameterization. It is possible to redefine scalars
calling function $\m{Vsetscalars[...]}$, or to add other scalars
calling function $\m{Vaddscalars[...]}$.

$\m{Vsetintvar[r]}$ - declares the symbol $\m{r}$ as integration
variable over which Feynman parameterization and automatic
$d$-dimensional integration will be performed.

$\m{Vsetsortlist[p1, \{mu1\}, \{mu2\}, p2, p3...]}$ - sorts
momenta ($\m{p1, p2, p3...}$) and $\gamma$-matrixes ($\m{\{mu1\},
\{mu2\}...}$) in the designated sequence.

$\m{Vcalc[exp, options]}$ - carries out calculations on expression
$\m{exp}$ and groups (sorts) result. The following may be
specified as control parameters ($\m{options}$): integration -
On/Off (by default it is switched off), a symbol of integration
($\m{integrating\to True/False,\, integrating\to r}$) (the latter
also implies $\m{integration\to True}$), Feynman parameterization
On/Off (by default it is switched off) ($\m{dofeynman\to
True/False}$), sorting On/Off, sorting order ($\m{sorting\to
True/False,\, sorting\to\{p, \{mu\}, k, \{nu\}, \{sigma\}\}}$)
(the latter also implies $\m{sorting\to True}$), rearrangement of
"edge"-momenta On/Off (by default it is switched off) ($\m{pps\to
True/False}$), calculation of a trace On/Off. (by default it is
switched off) ($\m{spur\to True/False}$)...., etc.

$\m{Vfeynman[x]}$ - performs Feynman parameterization of
expression $\m{x}$. Prior to calling of this function it is
necessary to define the variable of integration by calling
$\m{Vsetintvar[r]}$, with respect to which function
$\m{Vfeynman[x]}$ should make parameterization. Gives the output
in the following form: \{$K$, $scalar_1^{i_1}
scalar_2^{i_2}...scalar_n^{i_n}$, $int\_var\_0$, $power$, $delta$,
$N$\}, where $K$ - is the factor arising in a result of
parameterization, $scalar_1^{i_1}$, $scalar_2^{i_2}$, …
$scalar_n^{i_n}$ - are additional multipliers, if any arise,
$int\_var\_0$ - is the displacement of the integration variable,
i.e. in initial expression the integration variable (let it be $X$
for example) should be replaced by $X-int\_var\_0$, $power$ - is
the power of the resulting denominator after parameterization,
$delta$ - is the parameter of integration, $N$ - is not factored
part, i.e. multipliers and factors not dependent on the
integration variable, which have not been parameterizated
\cite{Peskin:1996}:
\begin{eqnarray}
\frac{1}{D1\,D2\,D3\,D4^\varepsilon}=K
N\int\frac{\mathrm{d}u1\mathrm{d}u2...\mathrm{d}un\,\,u1^{i1}...un^{in}\delta(1-u1-...-un)}
{[(\mathrm{int\_var}-\mathrm{int\_var\_0})^2+\mathrm{delta}]^{\mathrm{power}}}
\label{eq:feynparam}
\end{eqnarray}

When specifying Feynman parameterization option in function
$\m{Vcalc}$, all steps and substitutions are carried out
automatically (see expressions (\ref{eq:mathM1},
\ref{eq:mathM2})).

$\m{Vsetrules[\{p1, p2, c1\}, \{p1, c2\}...]}$ - defines rules of
scalar product, corresponding to - $((p1, p2)=c1, (p1,p1)=p1^2=c2
…$).

\section{Calculation of an expression by means of MatrixExp package}

Below we present an example of calculation with use of MatrixExp
package. It is calculated a Feynman diagram for process $b\to
s\gamma g$ \cite{H:66} in case of local operators $O_1,O_2$,
having the following form: $O_1=(\bar s_{L}\gamma_\mu T^a
c_{L})(\bar c_{L}\gamma^\mu T^a b_{L})$, $O_2=(\bar
s_{L}\gamma_\mu c_{L})(\bar c_{L}\gamma^\mu b_{L})$. On Fig.~2 we
present the two Feynman diagrams sum of which in \cite{H:66} is
denoted as $J_{\alpha\beta}$. Here $q$ designates photon momentum,
$r$ - gluon momentum.

\begin{figure}[h]
\vspace{0cm} \hspace{0cm}
\includegraphics[width=14cm,height=4cm]{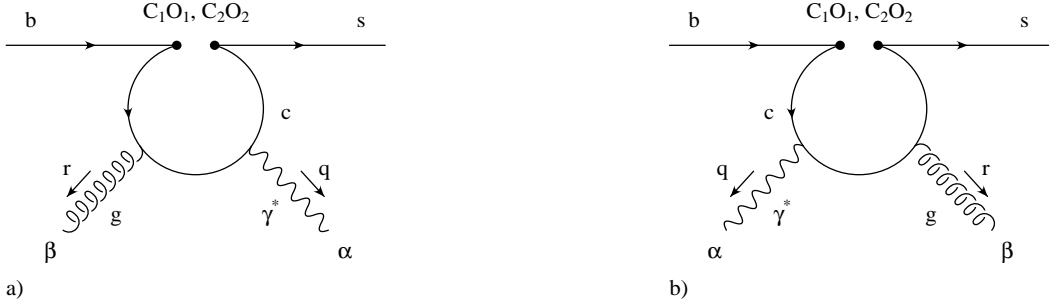}
\vspace{0cm} \caption[]{ Bremsstrahlung diagrams induced by  $O_1$
and $O_2$.}\vspace{0cm}
\end{figure}

\begin{eqnarray}
M_1&\sim&\bar{s}_L\frac{\gamma_\mu,\hat{k}+\hat{q}+m_c,\gamma_\alpha,\hat{k}+m_c,\gamma_\beta,
\hat{k}-\hat{r}+m_c,\gamma_\mu}{[(\hat{k}+\hat{q})^2-m_c^2][(\hat{k}-\hat{r})^2-m_c^2][k^2-m_c^2]}b_L,
\label{eq:m1}
\end{eqnarray}
\begin{eqnarray}
M_2&\sim&\bar{s}_L\frac{\gamma_\mu,\hat{k}+\hat{r}+m_c,\gamma_\beta,\hat{k}+m_c,\gamma_
\alpha,\hat{k}-\hat{q}+m_c,\gamma_\mu}{[(\hat{k}+\hat{r})^2-m_c^2][(\hat{k}-\hat{q})^2-m_c^2][k^2-m_c^2]}b_L,
\label{eq:m2}
\end{eqnarray}
\begin{eqnarray}
J_{\alpha\beta}&=&M_1+M_2.
\label{eq:jabdef1}
\end{eqnarray}
On start-up MatrixExp package by default defines some parameters,
in particular variables $u1... u9$ are defined as scalars,
automatic sorting is switched On, automatic integration is
switched Off.

First of all we add $mc$ as scalar
\begin{equation}
\m{Vaddscalars[mc]} \label{eq:mathmc1}
\end{equation}

Next we define the rules for scalar products $q^2=0$, $r^2=0$
\begin{equation}
\m{Vsetrules[\{q, 0\}, \{r, 0\}]} \label{eq:vsetrules1}
\end{equation}

Finally we call function $\m{Vcalc}$ for calculation of $M_1$ and
$M_2$ accordingly
\begin{eqnarray}\nonumber
\hspace{-0.5cm}
\m{M1}&=&\m{Vcalc}[\frac{\m{G[l1,R,\{mu\},k+q+mc,\{al\},k+mc,\{bet\},k-r+mc,\{mu\},L]}}{\m{((k-r)^2-mc^2)((k+q)^2-mc^2)(k^2-mc^2)}},\\
&&\m{integrating\to k, dofeynman\to True,
sorting\to\{\{al\},\{bet\},r,q\}}] \label{eq:mathM1}
\end{eqnarray}
\begin{eqnarray}\nonumber
\hspace{-0.5cm}
\m{M2}&=&\m{Vcalc}[\frac{\m{G[l1,R,\{mu\},k+r+mc,\{bet\},k+mc,\{al\},k-q+mc,\{mu\},L]}}{\m{((k+r)^2-mc^2)((k-q)^2-mc^2)(k^2-mc^2)}},\\
&&\m{integrating\to k, dofeynman\to True,
sorting\to\{\{al\},\{bet\},r,q\}}] \label{eq:mathM2}
\end{eqnarray}
where the option $\m{dofeynman\to True}$ turns On Feynman
parameterization, and the option $\m{integrating\to k}$ defines
the integration variable (and turns On integration). Received
$\m{M1}$ and $\m{M2}$ are grouped in sorted resulting matrix
expressions (e.g. $\m{G[l1, \{al\}, \{bet\}, q]}$, $\m{G[l1,
\{al\}, \{bet\}, r]}$, $\m{G[l1, \{al\}, \{bet\}]}$, $\m{G[l1,
\{bet\}, r, q]...}$) and include scalars $u1$, $u2$ and $u3$,
which arose due to Feynman parameterization. (actually one of the
scalars will be missing (in this case - $u1$) since during Feynman
parameterization, MatrixExp package automatically carries out
integration ($\int X\cdot\delta(1-u1-u2-u3)\mathrm{d}u1$),
replacing $u1$ by $1-u2-u3$). After that we consider the sum
$J_{\alpha\beta}=M1+M2$
\begin{equation} \m{Jab=Vcalc[M1+M2, sorting\to \{\{al\},
\{bet\}, r, q\}}] \label{Jab} \end{equation} or, using the built
in function of MATHEMATICA (since $\m{M1}$ and $\m{M2}$ are
already sorted in the identical order) \begin{equation}
\m{Jab=Simplify[M1+M2]} \label{Jab1}
\end{equation}

In the further we need to simplify expression (\ref{Jab},
\ref{Jab1}), using transversality of photon and gluon, Ward
identities, etc., and after representing through
$E[\alpha,\beta,r]$ we receive the form presented in \cite{H:66}.

\subsection*{ Acknowledgements}

The work was partially supported by the ANSEF-05-PS-hepth-813-100
program.

\newpage
\hspace{1pc}
{\bf TEST RUN OUTPUT}

\bigskip
In the package $d=4-2\varepsilon$ denotes the space-dimension, so
avoid its usage as momenta.\footnote{The last row "Timing: " shows
the tested time of the calculation. Results for Pentium IV 2.8GHz.
}
\begin{eqnarray}\nonumber
&&\mathin{361}{Vcalc[G[l1,a,\{mu\},b,\{mu\},a]]}\\
&&\mathout{361}{(-2+d)\,G[l1,b]\,a.a-2\,(-2+d)\,G[l1,a]\,a.b}\label{Out361}\\
\timing{0.05}
\end{eqnarray}
To be able to use scalars in expressions we must declare scalars,
otherwise by default they will be treated as momenta
\begin{eqnarray}\nonumber
&&\mathin{362}{Vaddscalars[ma,mb,mc]}\\
&&\mathout{362}{\{u1,u2,u3,u4,u5,u6,u7,u8,u9,ma,mb,mc\}}\label{Out362}\\
\timing{0.00}
\end{eqnarray}
We can use either global sorting (via $\m{Vsorton[True]}$ and
$\m{Vsetsortlist[a, b, c]}$, which will be automatically the
default sorting for all expressions) or local sorting
\begin{eqnarray}\nonumber
&&\mathin{363}{r0=Vcalc[G[l1,R,a+ma,b,c+mc,L]]}\\
&&\mathout{363}{ma\,mc\,G[l1,b,L]+G[l1,a,b,c,L]}\label{Out363}\\
\timing{0.00}
\end{eqnarray}
\begin{eqnarray}\nonumber
&&\mathin{364}{r1=Vcalc[G[l1,R,a+ma,b,c+mc,L],}\\
\nonumber&&\;\;\;\;\;\;\;\;\;\;\;\;\;\;\;\;\;\;\;\m{sorting\to \{c,b,a\}]}\\
\nonumber&&\mathout{364}{-G[l1,c,b,a,L]+2\,G[l1,c,L]\,a.b\,+}\\
&&\;\;\;\;\;\;\;\;\;\;\;\;\;\;\;\;\;\;\;\mathrm{G[l1,b,L]\,(ma\,mc-2\,a.c)+2\,G[l1,a,L]\,b.c}\label{Out364}\\
\timing{0.02}
\end{eqnarray}
\begin{eqnarray}\nonumber
&&\mathin{365}{r2=Vcalc[r1,sorting\to \{b,c,a\}]}\\
\nonumber&&\mathout{365}{G[l1,b,c,a,L]+G[l1,c,L]\,a.b+}\\
&&\;\;\;\;\;\;\;\;\;\;\;\;\;\;\;\;\;\;\;\mathrm{G[l1,b,L]\,(ma\,mc-2\,a.c)}\label{Out365}\\
\timing{0.00}
\end{eqnarray}
\begin{eqnarray}\nonumber
&&\mathin{366}{r3=Vcalc[r2,sorting\to \{a,b,c\}]}\\
&&\mathout{366}{ma\,mc\,G[l1,b,L]+G[l1,a,b,c,L]}\label{Out366}\\
\timing{0.02}
\end{eqnarray}
Here $L$ and $R$ denote $(1-\gamma_5)/2$ and $(1+\gamma_5)/2$
respectively. As we see, the Output (including numerical
coefficients and scalars) may be used as Input without any
additional transformations (e.g. $r1$, which is Out[364] is used
as input in (\ref{Out365})).

To use integration and Feynman parameterization, lets define some
rules for scalar product (we use arbitrary expressions and
definitions here, just as mathematical expressions)

\begin{eqnarray}\nonumber
&&\mathin{367}{Vsetrules[\{a,ma^2\},\{b,mb^2\},\{a,b,ma*mb/2\}]}\\
&&\mathout{367}{\{\{a,ma^2\},\{b,mb^2\},\{a,b,\frac{ma\,mb}{2}\}\}}\label{Out367}\\
\timing{0.02}
\end{eqnarray}

\begin{eqnarray}\nonumber
&&\mathin{368}{res=Vcalc[\frac{G[l1,R,a,k,\{mu\},b+mb,k,\{mu\},L}{((k-a)^2-ma^2)((k-b)^2-mb^2)},}\\
\nonumber&&\;\;\;\;\;\;\;\;\;\;\;\;\;\;\;\;\;\;\;\m{integrating\to k,dofeynman\to True]}\\
\nonumber&&\mathout{368}{-((-2+d)mb(4\pi)^{-2+\mathrm{eps}}}\\
\nonumber&&\;\;\;\;\;\;\;\;\;\;\;\;\;\;\;\;\;\;\;\mathrm{\left(-\frac{1}{-ma^2(-1+u2)^2+ma\,mb\,(-1+u2)-mb^2u2^2}\right)^{-1+\mathrm{eps}}}\\
\nonumber&&\;\;\;\;\;\;\;\;\;\;\;\;\;\;\;\;\;\;\;\mathrm{(-(-2+eps)\,(-ma^2(-1+u2)^2+ma\,mb\,(-1+u2)\,u2-}\\
\nonumber&&\;\;\;\;\;\;\;\;\;\;\;\;\;\;\;\;\;\;\;\mathrm{mb^2\,u2^2))+(-1+eps)\,(ma^2\,(-1+u2)^2+}\\
\nonumber&&\;\;\;\;\;\;\;\;\;\;\;\;\;\;\;\;\;\;\;\mathrm{ma\,mb\,(-1+u2)\,u2)-mb^2\,u2^2)+(-1+eps)}\\
\nonumber&&\;\;\;\;\;\;\;\;\;\;\;\;\;\;\;\;\;\;\;\mathrm{(ma^2(-1+u2)^2-ma\,mb\,(-1+u2)\,u2+mb^2\,u2^2))}\\
\nonumber&&\;\;\;\;\;\;\;\;\;\;\;\;\;\;\;\;\;\;\;\mathrm{G[l1,a,L]\,Gamma[-1+eps])/(-ma^2\,(-1+u2)^2+}\\
&&\;\;\;\;\;\;\;\;\;\;\;\;\;\;\;\;\;\;\;\mathrm{ma\,mb\,(-1+u2)\,u2-mb^2\,u2^2)}\label{Out368}\\
\timing{0.13}
\end{eqnarray}

We see, that all the substitutions after parameterization are
done, a Dirac-delta integration over parameter $u1$ is done,
integration over $k$ is done, and we have a final result.

Here we calculate the spur in two ways. First we calculate the
expression $r0$, followed by calculation of the spur of the
result, and then we directly calculate the spur of the initial
expression $r0$.

\begin{eqnarray}\nonumber
&&\mathin{369}{r0=G[l1,a+ma,b+mb,a+ma,c+mc,f+ma,L];}\\
\nonumber&&\;\;\;\;\;\;\;\;\;\;\;\;\;\;\;\;\;\;\;\m{rr=Vcalc[r0]}\\
\nonumber&&\;\;\;\;\;\;\;\;\;\;\;\;\;\;\;\;\;\;\;\m{Vcalc[rr,spur\to True]}\\
\nonumber&&\;\;\;\;\;\;\;\;\;\;\;\;\;\;\;\;\;\;\;\m{Vcalc[r0,spur\to True]}\\
\nonumber&&\mathout{370}{3\,ma^3\,mb\,mc\,G[l1,L]+3\,ma^2\,mb\,mc\,G[l1,a,L]+}\\
\nonumber&&\;\;\;\;\;\;\;\;\;\;\;\;\;\;\;\;\;\;\;\mathrm{3\,ma^3\,mb\,G[l1,c,L]+3\,ma^2\,mb\,mc\,G[l1,f,L]+}\\
\nonumber&&\;\;\;\;\;\;\;\;\;\;\;\;\;\;\;\;\;\;\;\mathrm{3\,ma^2\,mb\,G[l1,a,c,L]+3\,ma\,mb\,mc\,G[l1,a,f,L]+}\\
\nonumber&&\;\;\;\;\;\;\;\;\;\;\;\;\;\;\;\;\;\;\;\mathrm{3\,ma^2\,mb\,G[l1,c,f,L]+3\,ma\,mb\,G[l1,a,c,f,L]}\\
\nonumber&&\mathout{371}{6\,ma\,mb\,(ma\,a.c+mc\,a.f+ma\,(ma\,mc+c.f))}\\
&&\mathout{372}{6\,ma\,mb\,(ma\,a.c+mc\,a.f+ma\,(ma\,mc+c.f))}
\label{Out369}\\
\timing{0.09}
\end{eqnarray}

And finally here are the timings for calculations of
(\ref{eq:mathM1}--\ref{Jab}): {\bf 1.86}, {\bf 1.64} and {\bf
6.00} seconds respectively.

\end{document}